# Eye-Movement Control During the Reading of Chinese: An Analysis Using the Landolt-*C* Paradigm

Yanping Liu[1], Erik D. Reichle[2], & Ren Huang[3]

[1]Key Laboratory of Behavioral Science

Institute of Psychology, Chinese Academy of Sciences

[2]School of Psychology, University of Southampton, UK

[3]Department of Psychology, Sun Yat-Sen University, China

Author note

This research was supported by two grants from the China Postdoctoral Science Foundation (2013M541073 & 2014T70132) awarded to the first author, and by a Grant HD075800 from the National Institutes of Health awarded to the second author. Correspondence should be addressed to Yanping Liu, 16 Lincui Road, Key Laboratory of Behavioral Science, Institute of Psychology, Chinese Academy of Sciences, Beijing, China. Email: liuyp@psych.ac.cn.

**Abstract**

Participants in an eye-movement experiment performed a modified version of the Landolt-*C* paradigm (Williams & Pollatsek, 2007) in which they searched for target squares embedded in linear arrays of spatially contiguous "words" (i.e., short sequences of squares having missing segments of variable size and orientation). Although the distributions of single- and first-of-multiple fixation locations replicated previous patterns suggesting saccade targeting (e.g., Yan, Kliegl, Richter, Nuthmann, & Shu, 2010), the distribution of all forward fixation locations was uniform, suggesting the absence of specific saccade targets. Furthermore, properties of the "words" (e.g., gap size) also influenced fixation durations and forward saccade length, suggesting that on-going processing affects decisions about when and where (i.e., how far) to move the eyes. The theoretical implications of these results for existing and future accounts of eye-movement control are discussed.

Neural systems are extremely adept at exploiting the regularities that exist in the environment for the purposes of making (near) optimal inferences to guide behavior (Anderson, 1990). For example, the systems that mediate vision exploit regularities of objects (e.g., the Gestalt principle of similarity) to represent those objects in a manner that affords their invariant perception despite variability of both the objects and viewing angle, partial occlusion of the objects, etc. (Smith, 1988). It should therefore come as little surprise that complex visual tasks like reading also exploit regularity, allowing the task to be performed in a (near) optimal manner (Liu & Reichle, 2010; Liu, Reichle, & Gao, 2013; Reichle & Laurent, 2006; see also Bicknell & Levy, 2012). For example, because low-frequency words require more time to identify than high-frequency words (Forster & Chambers, 1973; Schilling, Rayner, & Chumbley, 1998), readers of alphabetic languages like English tend to spend more time fixating the low- than high-frequency words (Inhoff & Rayner, 1986; Kliegl, Nuthmann, & Engbert, 2006; Rayner, Ashby, Pollatsek, & Reichle, 2004). And in a similar manner, because words can be identified most rapidly when they are fixated near their center (O'Regan, 1981; Rayner, 1979; Rayner & Morrison, 1981), readers of alphabetic languages use the blank spaces between words to direct their eyes to locations near the center of words (McConkie, Kerr, Reddix, & Zola, 1988; Rayner, Sereno, & Raney, 1996). Such eye-movement behavior takes advantage of the inherent regularities of the text to support reading that is rapid, but that also maintains some overall level of comprehension. What is less clear, however, is whether such behaviors are universal, generalizing across languages and writing systems that exhibit less—or perhaps different—patterns of regularity. One prime example that has been the focus of much recent research is Chinese (for a review, see Zang, Liversedge, Bai, & Yan, 2011).

The Chinese language and writing system has many properties that make it different from the alphabetical languages that have most often been used in experiments to understand reading. For example, in contrast to alphabetic writing systems, Chinese words are not comprised of letters and are not spatially segregated by blank spaces. Instead, Chinese words consist of 1-4 equally sized box-shaped characters that are comprised of 1-36 "strokes" (i.e., line segments that originally corresponding to the brush strokes used to write the characters), with the characters being arranged in a spatially adjacent linear array (see Figure 1 for an example). The fact that Chinese words are comprised of a variable number of characters and not demarcated by blank spaces means that Chinese readers must use their knowledge of their language to somehow segment character sequences into their corresponding words (e.g., see Li, Rayner, & Cave, 2009). This task of segmenting characters into words is intrinsically difficult because there is often ambiguity in how a character sequences might be segmented, with a given sequence of four characters, for example, possibly corresponding to two 2-character words or one 4-character word (Hoosain, 1991, 1992; Liu, Li, Lin, & Li, 2013).

----- Figure 1 -----

Because of these properties of written Chinese words, there is some debate about both the extent to which and how the eye movements of Chinese readers are influenced by the lexical properties of words (as they clearly are in alphabetic languages; see Rayner, 1998, 2009). For example, one theoretical "camp" has argued that characters are more important than words during the reading of Chinese, and that characters are the basic units that determine both when the eyes move (Chen, 1996; Chen, Song, Lau, Wong, & Tang, 2003; Chen & Zhou, 1999; Feng, 2008; Hoosain, 1991, 1992) and where the eyes move (Tsai & McConkie, 2003; Yang & McConkie, 1999). In contrast, the other theoretical "camp" maintains that, because words have a psychological reality in the minds of Chinese readers (Cheng, 1981; Li et al., 2009; Li, Bicknell, Liu, Wei, & Rayner, 2014; Li & Logan, 2008; Li, Gu, Liu, & Rayner, 2013), they also play an important functional role in determining when the eyes move during the reading of Chinese (Liversedge, Hyönä, & Rayner, 2013; Yan, Tian, Bai, & Rayner, 2006). However, this second perspective further diverges in their accounts of how readers of Chinese decide where to move their eyes. One contingent has argued that the locations of initial fixations on words are a function of whether the words are successfully segmented in the parafovea, with saccades towards segmented words being directed towards their centers but saccades towards un-segmented words being directed towards their beginnings (Yan et al., 2010). In contract, the other contingent has proposed that there is no "default" saccade target within a word, but that the length of the saccade moving the eyes into a word is instead influenced by the relative difficulty of foveal and/or parafoveal processing, with easier processing resulting in longer forward saccades (Liu, Reichle, & Li, 2014; Wei, Li, & Pollatsek, 2013). Given these different theoretical perspectives on eye-movement control during the reading of Chinese, it is clear that—in contrast to alphabetic languages—we still know very little about the basic mechanisms that determine when and where the eyes move during the reading of non-alphabetic languages, or more specifically, about *if* and, if so, *how* lexical processing mediates eye-movement control during the reading of Chinese.

One strategy that has been used to examine the influence of lexical processing in guiding eye movements during the reading of alphabetic languages has involved comparisons of such eye movements to those that are observed during tasks that do not require lexical processing but that do entail many of the same visual and oculomotor constraints, such as the Landolt-*C* "reading" task (Corbic, Glover, &

Radach, 2007; Williams & Pollatsek, 2007; Vanyukov, Warren, Wheeler, & Reichle, 2012; see also Williams, Pollatsek, & Reichle, 2014). In experiments that have used this task, participants were instructed to scan through linear arrays of Landolt *C*s (i.e., ring-shaped stimuli having missing segments (i.e., gaps) of variable size and/or orientations) to search for target stimuli—rings containing no missing gaps. Because these linear arrays are comprised of Landolt-*C* stimuli arranged into sequences of "words" (i.e., clusters of the stimuli separated by blank spaces between clusters), the task has been used to study saccadic targeting under conditions that resemble those experienced during the reading of alphabetic languages, minus all of the language-related behaviors that are normally engaged during word, sentence, and discourse processing. The key findings from these experiments is that participants seem to treat the Landolt-*C* clusters as "words", directing their eyes towards their centers and fixating the easier-to-process clusters (i.e., those with larger gaps sizes and/or that occur more often during the course of the experiment) for shorter durations than the more difficult-to-process clusters (i.e., those with smaller gap sizes and/or that occur less often during the experiment).

In the experiment that is reported in this article, the Landolt-*C* task was modified so that it more closely resembled the reading of Chinese (see Figure 2), thereby providing a novel way to examine if the segmentation of "words" (i.e., spatially adjacent stimuli having similarly sized and orientated gaps) affects fixation durations and/or saccadic targeting[1]. That is, our participants were instructed to scan linear arrays of Landolt-squares, or boxes with a missing segment of varying size and orientation, and to indicate the number of targets (i.e., squares that do not contain missing segments; e.g., □). The advantages of this method are that it provides a way of precisely manipulating both word processing difficulty (i.e., gap size) and size (i.e., the number of identical spatially adjacent non-target squares) in the absence of language processing.

The main goal of this experiment was therefore to determine if/how word segmentation and identification affects eye movements in this task so as to better inform our understanding of the word-based effects that have been observed during the actual reading of Chinese text. More specifically, we focus on: (1) if/how fixation durations on individual words are affected by their identification difficulty; and (2) if/how saccadic targeting is affected by their segmentation difficulty. If easier-to-identify words are fixated for shorter durations but saccade targeting is unaffected by the ease of word segmentation, then this pattern of results would lend further support to the hypothesis that the processing associated with words influences both when and where the eyes move during the reading of Chinese (e.g., Liu et al., 2014; Wei et al., 2013), but that parafoveal word segmentation is not the primary determinant of where the eyes are moved (cf., Yan et al., 2010).

## Method

### Participants

Twenty participants (12 males) recruited from Sun Yat-Sen University were paid for their participation. All participants had normal or corrected-to-normal vision and were naive to the purpose of the experiment.

### Apparatus

Stimuli were displayed on a 23-in. LCD monitor (Samsung SyncMaster2233) using SR-Research Experiment Builder software. The monitor had a resolution of $1,680 \times 1,050$ pixels and refresh rate of 120 Hz. Eye movements were recorded using a SR-Research Eyelink 1000 eye tracker (Kanata, ON, Canada) with a spatial resolution of 0.01° and sampling rate of 1,000 Hz.

**Stimuli and Design**

The materials were designed to be as similar to Chinese sentences as possible (e.g., cf., Figures 1 vs. 2). Each "character" was a Landolt-Square (40 pixels × 40 pixels) with a gap that was 2, 4, 6, or 8 pixels in size and that occurred in the left, right, top, or bottom of the character. Each cluster or "word" was composed of 1, 2, 3, or 4 characters. Both the gap size and location were held constant within a given word, but varied between words. The spaces between any two successive characters were 6 pixels. Each array of words or "sentence" contained 10 randomly selected words and ranged from 16 to 33 characters in length ($M = 25$ characters). The combination of the number of characters per word and the gap size and orientation in a given word gave 64 unique words. Each word was repeated 40 times across the experiment. Each sentence had 0, 1, or 2 targets (i.e., characters without gaps) that could appear with equal probability within any word except the first and last word within a sentence. Participants were instructed to scan through each sentence and then indicate via button presses how many targets occurred in the sentence, with one sentence being displayed per trial. There were four blocks containing 64 trials each, with the order of blocks being completely random.

**----- Figure 2 -----**

**Procedure**

Upon arrival, participants were given task instructions and then seated 63 cm from the video monitor. A chinrest was used to minimize head movements during the experiment. Although viewing was binocular, eye-movement data were only collected from right eye. The eye-tracker was calibrated and validated using a 9-dot procedure at the beginning of each block of trials, with additional calibrations and validations being conducted as necessary. After that, participants completed 8 practice trials (not included in our analyses) and then completed the 4 blocks of experimental trials. A drift-check procedure was performed before each trial; a sentence was displayed after the participants successfully fixated the white box ($1° \times 1°$) at the location of the first character in the sentence. Participants were instructed to view a right-bottom region of the screen to terminate a trial and to then press the number button on the keyboard corresponding to the number of targets that appeared in the trial. Accuracy feedback was randomly provided on half of the trials.

## Results

### Accuracy

The mean overall accuracy of the experiment was 91% ($SD = 0.05$).

### Eye-Movement Results

Fixations on the first and last words in each sentence were removed from our analyses because the former coincided with the abrupt appearance of the sentence and the latter coincided with the termination of a trial. Fixations on words containing targets and immediately preceding or following targets were also removed from our analyses. Thus, fixations on 64.3% of the total number of words were included in our analyses. From these data, saccades longer than 3 $SD$s above the mean for a given participant were excluded (3.05% of the total saccades).

To determine if/how the processing of words affected eye movements during our task, our analyses focused on the relationship between various eye-movement dependent measures (related to when and where to move eyes) and properties of both the fixated word (i.e., word *N*) and the two spatially adjacent words (i.e., words *N*-1 and *N*+1). To do this, we examined the first-pass fixation-location distributions corresponding to: (1) first fixations; (2) single fixations; (3) first-of-multiple fixations; and (4) all forward fixations. We also examined the following measures: (5) *forward-saccade length*, or the distance between two consecutive forward fixations during first-pass scanning; (6) *first-fixation duration*, or the duration of the first fixation on a word during first-pass scanning; (7) *gaze duration*, or the sum of all first-pass fixation durations on a word; and (8) *total-viewing time*, or the sum of all fixation durations on a word. We then built linear mixed-effects models (*LMM*s) for each measure, specifying participants, items, and random slopes of each predictor (e.g., word properties) as the random effects so that our reported significance values reflect the variability of participants, items, and the slopes of the predictors (see also Barr, Levy, Scheepers, & Tily, 2013). Based on prior results and our priori hypotheses, we also included practice (i.e., ordinal trial number) and the properties of words *N*-1, *N*, and *N*+1 (i.e., gap size and number of characters) as fixed-effect factors in our LMMs. The models were then fitted using the lme4 package (ver. 1.1-7; Bates, Maechler, Bolker, & Walker, 2014; Pinheiro & Bates, 2000) and *p*-values were estimated using lmerTest package (ver. 2.0-20; Kuznetsova, Brockhoff, & Christensen, 2013) in *R* (ver. 3.1.2; R Development Core Team, 2015).

**Fixation Locations**

To determine if parafoveal word segmentation influences saccadic targeting, we analyzed our various measures of fixation location during first-pass scanning. Figure 3 respectively shows the distributions of initial-, single-, first-of-multiple-, and all forward fixation locations as a function of the number of characters with the words, their gap sizes, and practice (i.e., ordinal block number). As shown, the locations of initial fixations (panels a-b) and the first-of-multiple fixations (panels c-d) were more likely to be near the beginnings of words, whereas single fixations (panels e-f) were more likely to be located near the centers of words. However, the locations of all forward fixations (panels g-h) were uniform (all values of $\chi^2 \leq 1.74$, $p$s $\geq 0.755$; also see Table 1). All of these results are consistent with previous findings that have been observed with Chinese reading (Li, Liu, & Rayner, 2011; Yan et al., 2010).

**----- Figure 3 & Table 1 -----**

LMMs were further used to examine if practice and the properties of words $N$-1, $N$, and $N$+1 affected all forward fixation locations on word $N$ (measured from the left edge of that word). Unsurprisingly, Table 2 shows that the mean fixation location across all forward saccades moved further to the right of word $N$ as its length increased ($b = 0.99$, $SE = 0.01$, $t = 97.27$, $p < 0.001$). The fact that the mean fixation location was not influenced by the gap size of word $N$, the properties of the spatially adjacent words, or the amount of practice (all values of $|t| \leq 1.24$, all $p$s $\geq 0.214$) indicates that the uniformity of distributions of forward fixation locations was robust, which in turn suggests the absence of any (strongly) preferred viewing location on the words in our task. Importantly, the "preferred viewing locations" which were seemingly evident in the distributions of first and first-of-multiple fixations (i.e., located near the word beginnings) and of the single fixations (i.e., located near the word centers) were specious because of the arbitrary way in which these fixations were extracted from the uniform distribution of all forward fixations. We will say more about this in the Discussion.

**----- Table 2 -----**

**Saccade Length**

Although the results from fixation-location analyses indicated that properties of neither the fixated word nor its spatially adjacent words influenced the distribution of all of the forward fixation locations, it is possible that foveal and/or parafoveal processing influenced the length of forward saccade. As Table 3 indicates, forward saccade length increased with the lengths of word $N$-1 ($b = 0.01$, $SE = 0.01$, $t = 2.43$, $p = 0.015$), word $N$ ($b = 0.10$, $SE = 0.01$, $t = 9.88$, $p < 0.001$), and word $N$+1 ($b = 0.04$, $SE = 0.01$, $t = 5.05$, $p < 0.001$). Forward saccade length also increased with the gap size of word $N$-1 ($b = 0.01$, $SE = 0.003$, $t = 1.72$, $p = 0.085$), word $N$ ($b = 0.01$, $SE = 0.01$, $t = 2.23$, $p = 0.037$), and word $N$+1 ($b = 0.02$, $SE = 0.003$, $t = 5.58$, $p < 0.001$). These results were entirely consistent with those of previously reported experiments involving Chinese reading which demonstrated that forward saccade length increases with both the processing ease (frequency) and length of the fixated word (Li et al., 2014; Liu et al., 2014; Liu, Li, &Pollatsek, submitted for review; Wei et al., 2013), and with the processing ease (frequency) and length of upcoming (parafoveal) word (Li et al., 2014; Liu et al., 2014; Liu et al., submitted for review).

----- Table 3 -----

**Fixation Durations**

Our analyses have shown that properties of the fixated words and its spatially adjacent words did not influence the uniformity of all forward fixation location distributions, but that word properties did influence the length of forward saccades. This suggests that ongoing processing influences saccade length rather than fixation locations on words. It is therefore also necessary to examine how word properties and practice affected ongoing processing.

As Table 4 shows, first-fixation durations decreased with the length of word $N$ ($b = -1.35$, $SE = 0.67$, $t = -2.01$, $p = 0.051$) and the gap size of word $N$ ($b = -1.16$, $SE =$ 0.33, $t$ = -3.50, $p$ = 0.002). Conversely, as Table 5 shows, gaze durations increased with the length of word $N$ ($b$ = 135.70, $SE$ = 10.91, $t$ = 12.44, $p$ < 0.001), but decreased with the lengths of word $N$-1 ($b$ = -4.03, $SE$ = 1.21, $t$ = -3.34, $p$ < 0.001) and word $N$+1 ($b$ = -6.84, $SE$ = 1.53, $t$ = -4.46, $p$ < 0.001). Similarly, gaze durations also decreased with increasing gap size of word $N$ ($b$ = -9.75, $SE$ = 0.72, $t$ = -13.47, $p$ < 0.001).

##### ----- Table 4 & 5 -----

Finally, as Table 6 shows, total-viewing times (like gaze durations) increased with the length of word $N$ ($b$ = 158.70, $SE$ = 3.66, $t$ = 43.37, $p$ < 0.001), but decreased with the lengths of word $N$-1 ($b$ = -5.06, $SE$ = 1.41, $t$ = -3.59, $p$ < 0.001) and word $N$+1 ($b$ = -6.53, $SE$ = 1.41, $t$ = -4.62, $p$ < 0.001). And like first-fixation and gaze durations, the total-viewing times deceased with increasing gap size of word $N$ ($b$ = -13.16, $SE$ = 0.85, $t$ = -15.44, $p$ < 0.001).

##### ----- Table 6 -----

The results of the preceding analyses are consistent with other experimental results showing that the properties of the fixated word and its spatially adjacent neighbors can influence fixation durations during Chinese reading. For example, several studies have demonstrated that fixation durations decrease with increasing processing ease (e.g., frequency) of both the fixated word (Yan et al., 2006; Yang & McConkie, 1999; Rayner, Li, Juhasz, & Yan, 2005) and the previously fixated word (i.e., spillover effects; Li et al., 2014, see Table 1). Our results are also consistent with evidence that gaze durations tend to increase as the length of the fixated word increases and as the length of the previously fixated word decreases (Li et al., 2014, see Table 1). The theoretical implications of these results for our understanding of eye-movement control during the reading of Chinese will be discussed next.

## General Discussion

The present experiment aimed to determine *if* and *how* "word" processing affects eye movements in a Chinese-like visual search task so as to better inform our understanding of the word-related effects that have been observed in the actual reading of Chinese text. We replicated the well-established pattern of fixation locations that has been reported in Chinese reading: Whereas the distribution of first-fixation locations peaked at the word beginnings when the words were fixated more than once, the distribution of single-fixation locations peaked near the word centers (Li et al., 2011; Yan et al., 2010). However, an examination of the distributions of all of the forward fixation locations indicated that they were clearly uniform and not influenced by word length, word-processing difficulty (i.e., gap size), or practice. Finally, there were effects of word length and word-processing difficulty on the various fixation-duration measures and forward saccade length, suggesting that variables that influence that rate of on-going word processing also dynamically influences both when and where the eyes move during our task and—as we will argue, by extension—the reading of Chinese text (Liu et al., 2014). In the remainder of this discussion, we indicate precisely why we maintain this position, and discuss in more detail what we believe each of the preceding results says about how readers of Chinese make "decisions" about when and where to move their eyes.

First, the finding that the distributions of all forward fixation locations were uniform and invariant to both "word" properties and practice strongly suggests that there is no preferred viewing location in our task; otherwise, the distributions would have tended to be normal, peaked on whatever viewing location actually afforded efficient performance of our task. We believe that the patterns of fixation-location distributions observed for first-of-multiple and single fixations in our study and others (e.g., Li et al., 2011) and that have been suggested as indicative of a preferred viewing location that is conditional upon the successful segmentation of parafoveal words (e.g., see Yan et al., 2010) is instead an artifact of the simple fact that words that are fixated near their centers are less likely to be refixated than words that are fixated nears their beginnings (Liu et al., submitted for review). Li et al. (2011) provide convincing support for this conjecture by running simulations of eye movements during Chinese reading that included the simple assumption that saccades are of constant mean length but with some variability. These simulations reproduced the findings that the distributions of single- and first-of-multiple fixation locations tend to be near the centers and beginnings of words, respectively, thereby demonstrated precisely how a pattern of results that might otherwise suggest a type of saccade targeting that is conditional upon parafoveal word segmentation can emerge from simple assumptions that involve no specific saccade targets (see also Ma, Li, & Pollatsek, in press). These simulations, in conjunction with our findings, are therefore consistent with the hypothesis that there are no specific saccade targets during the reading of Chinese. In other words, a saccade target is not a true target per se, but is instead the consequence of a saccade that is intended to move the eyes to an informative viewing location.

Second, the finding that properties of words influenced the various fixation-duration measures in our task is completely consistent with findings that lexical properties of Chinese words also influence such measures during the reading of Chinese (Li et al., 2014; Liu et al., 2014; Liu et al., submitted for review; Yan et al., 2006; Yang & McConkie, 1999; Rayner et al., 2005). These findings indicate the validity of our method of using our version of the Landolt-*C* paradigm to examine eye-movement control during Chinese reading by replicating key results showing that the relative difficulty of "word" processing (Williams & Pollatsek, 2007; Williams et al., 2014) can influence the time spent fixating (processing) a given "word", analogous to what is known to happen in actual reading (e.g., Schilling et al., 1998). These findings are therefore important because they permits one to infer that whatever perceptual (e.g., segmenting individual "words" based on character gap size and orientation) and cognitive (e.g., discriminating non-targets from targets) are operative in our task are likely to have analogs in Chinese reading (e.g., segmenting individual words, identifying characters and words, etc.), and that, in both tasks, these processes are likely to play causal roles in the patterns of eye movements that are observed. The most transparent of these causal roles is undoubtedly related to the moment-by-moment "decisions" about when to move the eyes from one location to the next, but as our experiment demonstrates, the causal influence is also likely to extend to decisions about where to move the eyes.

This last conjecture is consistent with our third main finding that the properties of "words" influences forward saccade length in our task. On some level, this finding should not be too surprising if one considers the range on possible ways that readers might in theory guide their eyes while reading Chinese. On one end of the continuum, it is possible—thought not likely—that readers simply move their eyes randomly to new locations, perhaps adopting some simple heuristic (e.g., the "fixed saccade length" assumption used in the simulations reported by Li et al., 2011; also see Yan et al., 2010). At the other end of the continuum, readers might use word boundary information to move their eyes to the center of the next unidentified word, as posited to occur in models that simulate the eye movements of people reading alphabetic languages like English and German (e.g., *E-Z Reader*: Reichle, Pollatsek, Fisher, & Rayner, 1998; Reichle, Warren, & McConnell, 2009; Reichle, 2011; *SWIFT*: Engbert, Nuthmann, Richter, & Kliegl, 2005; Schad & Engbert, 2012). This second possibility also seems unlikely given the empirical evidence against preferred viewing locations in both this experiment and in experiments involving Chinese reading (e.g., Li et al., 2011), and given the inherent difficulties associated with word segmentation in Chinese reading (e.g., see Li et al., 2009). By the process of elimination, then, the fact that these two extreme possible accounts of saccade targeting in Chinese are not very feasible suggests a third possibility—that readers decide where to move their eyes using some type of information other than word boundaries. What might this information be?

One possibility that we believe is a viable candidate is whatever processing difficulty is associated with the currently fixated word and/or the word(s) immediately to the right of fixation. By this account, readers might use the *foveal load*, or the difficulty associated with identifying the fixated word, as a metric to gauge how far to move their eyes with the next saccade. The intuition behind this simple heuristic is that, because difficult-to-process fixated words afford less parafoveal preview of the upcoming word than do less difficult-to-process fixated words (Henderson & Ferriera, 1990; Kennison & Clifton, 1995), foveal load might provide the reader with a metric of how much processing still remains to be done on the parafoveal word when it is fixated, and thus a metric of how far the eyes should be moved into the word. (A parafoveal word that has received little processing during preview is likely to require a substantial amount of processing after it is fixated, making a more conservative fixation near the beginning of the word, thereby allowing for the possibility of one or more additional fixations on the word during the forward pass through the text.) And for the same reason, information that becomes available about the parafoveal word from the foveal word might also be expected to play an important role in modulating the length of the saccade leaving the foveated word if the goal is to move the eyes just past the location where the parafoveal characters and/or words have been sufficiently processed for their identification to be imminent. Such a proposal has already been suggested to account for experimental results showing that the saccade length leaving a word is modulated by the difficulty (e.g., frequency) of that word during Chinese reading (Liu et al., 2014; Wei et al., 2013), and as a general account of how people decide where to move their eyes during the reading of Chinese (Liu et al., submitted for review).

Thus, our variant Landolt-*C* paradigm has provided new information that is consistent with prior research using this paradigm (e.g., Williams & Pollatsek, 2007; Vanyukov et al., 2012) and with prior research on Chinese reading (e.g., Li et al., 2014; Liversedge et al., 2013; Yan et al., 2006), lending further support to a theoretical account of eye-movement control in Chinese reading—an account based on local character- and word-based control of *when* the eyes move that in turn affects *where* (or more precisely, how far) the eyes move (Liu et al., 2014, submitted for review; Wei et al., 2013). This represents significant empirical and theoretical progress in understanding eye-movement control during reading because it reveals the limitations of existing accounts of eye-movement control (e.g., the computational models cited earlier) that are based of years of research of alphabetic languages. It goes without saying that this progress will be extremely useful for developing formal accounts (i.e., computational models) of eye-movement control in non-alphabetic languages like Chinese, and more generally for informing our understanding of how differences among languages and writing systems influence the perceptual, cognitive, and motoric processes that move the eyes during reading.

# References

Anderson, J. R. (1990). The adaptive character of thought. Hillsdale, NJ: Erlbaum.

Barr, D. J., Levy, R., Scheepers, C., & Tily, H. J. (2013). Random effects structure for confirmatory hypothesis testing: Keep it maximal. *Journal of Memory and Language, 68*, 255-278.

Bates, D., Maechler, M., Bolker, B., & Walker, S. (2014). lme4: Linear mixed-effects models using Eigen and S4. URL http://lme4.r-forge.r-project.org/.

Bicknell, K., & Levy, R. (2012).The utility of modelling word identification from visual input within models of eye movements in reading. *Visual Cognition, 20*, 422-456.

Chen, H. (1996). Chinese reading and comprehension: A cognitive psychology perspective. In M. H. Bond (Ed.), *Handbook of Chinese psychology* (pp. 43-62). Hong Kong: Oxford University Press.

Chen, H., Song, H., Lau, W. Y., Wong, K. F. E., & Tang, S. L. (2003). Chinese reading and comprehension: A cognitive psychology perspective. In C. McBride-Chang & H. Chen (Eds.), *Reading development in Chinese children* (pp. 157-169). Westport, CT: Praeger.

Chen, H., & Zhou, X. (1999). Processing East Asian languages: An introduction. *Language and cognitive processes, 14*, 425-428.

Cheng, C. (1981). Perception of Chinese character. *Act Psychological Taiwanica, 23*, 137-153.

Corbic, D., Glover, L., & Radach, R. (2007). The Landolt-C string scanning task as a proxy for visuomotor processing in reading. A pilot study. In *Poster session presented at the 14th European Conference on Eye Movements*.

Engbert, R., Nuthmann, A., Richter, E., & Kliegl, R. (2005). SWIFT: A dynamical
model of saccade generation during reading. *Psychological Review, 112*, 777-813.

Feng, G. (2008). Orthography and eye movements: The paraorthographic linkage
hypothesis. In K. Rayner, D. Shen, X. Bai, & G. Yan (Eds.), *Cognitive and cultural influences on eye movements* (pp. 395-420). Tianjin, China: Tianjin People's Publishing House.

Forster, K. I. & Chambers, S. M. (1973). Lexical access and naming time. *Journal of Verbal Learning and Verbal Behavior, 12*, 627-635.

Henderson, J. M., & Ferreira, F. (1990). Effects of foveal processing difficulty on the
perceptual span in reading: Implications for attention and eye movement control. *Journal of Experimental Psychology: Learning, Memory, and Cognition, 16*, 417-429.

Hoosain, R. (1991). Aspects of the Chinese language. In R. Hoosain (Ed.),
*Psycholinguistic implications for linguistic relativity: A case study of Chinese* (pp. 5–21). Hillsdale, NJ: Erlbaum.

Hoosain, R. (1992). Psychological reality of the word in Chinese. In H. C. Chen & O.
J. L. Tzeng (Eds.), *Language processing in Chinese* (pp. 111-130). Amsterdam, the Netherlands: North-Holland.

Inhoff, A. W., & Rayner, K. (1986). Parafoveal word processing during eye fixations
in reading: Effects of word frequency. *Perception & Psychophysics, 40*, 431-439.

Kennison, S. M., & Clifton, C., Jr. (1995). Determinants of parafoveal preview benefit
in high and low working memory capacity readers: implications for eye movement control. *Journal of Experimental Psychology: Learning, Memory and Cognition, 21*, 68-81.

Kliegl, R., Nuthmann, A., & Engbert, R. (2006). Tracking the mind during reading: The influence of past, present, and future words on fixation durations. *Journal of Experimental Psychology: General, 135*, 12-35.

Kuznetsova, A., Brockhoff, P. B., & Christensen, R. H. B. (2013). lmerTest: Tests for random and fixed effects for linear mixed effect models (lmer objects of lme4 package). URL http://lmertest.r-forge.r-project.org/.

Li, X., Bicknell, K., Liu, P., Wei, W., & Rayner, K. (2014). Reading Is Fundamentally Similar Across Disparate Writing Systems: A Systematic Characterization of How Words and Characters Influence Eye Movements in Chinese Reading. *Journal of Experimental Psychology: General, 143*, 895-913.

Li, X., Gu, J., Liu, P., & Rayner, K. (2013). The advantage of word-based processing in Chinese reading: Evidence from eye movements. *Journal of Experimental Psychology: Learning, Memory, and Cognition, 39*, 879-889.

Li, X., Liu, P., & Rayner, K. (2011). Eye movement guidance in Chinese reading: Is there a preferred viewing location? *Vision Research, 51*, 1146-1156.

Li, X., & Logan, G. (2008). Object-based attention in Chinese readers of Chinese words: Beyond Gestalt principles. *Psychonomic Bulletin & Review, 15*, 945-949.

Li, X., Rayner, K., & Cave, K. R. (2009). On the segmentation of Chinese words during reading. *Cognitive Psychology, 58*, 525-552.

Liu, P., Li, W., Lin, N., & Li, X. (2013). Do Chinese readers follow the National Standard Rules for word segmentation during reading? *PLoS One, 8*, e55440.

Liu, Y., Li, X., & Pollatsek, A. (2014). Cognitive Control of Saccade Amplitude

    during the Reading of Chinese: A Theoretical Analysis and New Evidence.

    Manuscript submitted to review.

Liu, Y., & Reichle, E. D. (2010). The emergence of adaptive eye movements in

    reading. In S. Ohlsson & R. Catrabone (Eds.), *Proceedings of the 32nd Annual*

    *Conference of the Cognitive Science Society* (pp. 1136-1141). Austin, TX:

    Cognitive Science Society.

Liu, Y., Reichle, E. D., & Gao, D.-G. (2013). Using reinforcement learning to

    examine dynamic attention allocation during reading. *Cognitive Science, 37*,

    1507-1540.

Liu, Y., & Reichle, E. D., & Li, X. (2014). Parafoveal Processing Affects Outgoing

    Saccade Length During the Reading of Chinese. *Journal of Experimental*

    *Psychology: Learning, Memory, and Cognition.* Manuscript in press.

Liversedge, S.P., Hyönä, J., & Rayner, K. (2013). Eye movements during Chinese

    reading. *Journal of Research in Reading, 36*, S1-S3.

Ma, G., Li, X., & Pollatsek, A. (in press). There is no relationship between the

    preferred viewing location and word segmentation in Chinese reading. *Visual*

    *Cognition*. doi: 10.1080/13506285.2014.1002554.

McConkie, G. W., Kerr, P. W., Reddix, M. D., & Zola, D. (1988). Eye movement

    control during reading: I. The location of the initial eye fixations on words.

    *Vision Research, 28*, 1107-1118.

O'Regan, J. K. (1981). The convenient viewing position hypothesis. In D. F. Fisher,

    R. A. Monty, & J. W. Senders (Eds.), *Eye movements: Cognition and visual*

    *perception* (pp. 363–383). Hillsdale, NJ: Lawrence Erlbaum Associates.

Pinheiro, J.C., & Bates D.M. (2000). *Mixed-effects Models in S and S-PLUS*. Springer, New York, USA.

Rayner, K. (1979). Eye guidance in reading: Fixation locations within words. *Perception, 8*, 21-30.

Rayner, K. (1998). Eye movements in reading and information processing: 20 years of research. *Psychological Bulletin, 124,* 372-422.

Rayner, K. (2009). Eye movements and attention in reading, scene perception, and visual search. *Quarterly Journal of Experimental Psychology, 62*, 1457-1506.

Rayner, K., Ashby, J., Pollatsek, A., & Reichle, E. D. (2004). The effects of frequency and predictability on eye fixations in reading: Implications for the E-Z Reader model. *Journal of Experimental Psychology: Human Perception and Performance, 30*, 720-730.

Rayner, K., Li, X., Juhasz, B. J., & Yan, G. (2005). The effect of word predictability on the eye movements of Chinese readers. *Psychonomic Bulletin & Review, 12*, 1089-1093.

Rayner, K., & Morrison, R. E. (1981). Eye movements and identifying words in parafoveal vision. *Bulletin of the Psychonomic Society, 17*, 135-138.

Rayner, K., Sereno, S. C., & Raney, G. E. (1996). Eye movement control in reading: A comparison of two types of models. *Journal of Experimental Psychology: Human Perception and Performance, 22*, 1188-1200.

Reichle, E. D. (2011). Serial attention models of reading. In S. P. Liversedge, I. D. Gilchrist, & S. Everling (Eds.), *Oxford handbook on eye movements* (pp. 767-786). Oxford, U.K.: Oxford University Press.

Reichle, E. D. & Laurent, P. (2006). Using reinforcement learning to understand the emergence of "intelligent" eye-movement behavior during reading. *Psychological Review, 113*, 390-408.

Reichle, E. D., Pollatsek, A., Fisher, D. L., & Rayner, K. (1998). Toward a model of eye movement control in reading. *Psychological Review, 105*, 125-157.

Reichle, E. D., Warren, T., & McConnell, K. (2009). Using E-Z Reader to model the effects of higher-level language processing on eye movements during reading. *Psychonomic Bulletin & Review, 16*, 1-21.

Schad, D.J., & Engbert, R. (2012). The zoom lens of attention: Simulating shuffled versus normal text reading using the SWIFT model. *Visual Cognition, 20*, 391-421.

Schilling, H. E. H., Rayner, K., & Chumbley, J. I. (1998). Comparing naming, lexical decision, and eye fixation times: Word frequency effects and individual differences. *Memory and Cognition, 26*, 1270-1281.

Smith, B. (1988). Gestalt Theory: An Essay in Philosophy. In B. Smith (Ed.), *Foundations of Gestalt Theory* (pp. 11-81), Munich and Vienna: Philosophia Verlag.

Tsai, J. L., & McConkie, G. W. (2003). Where do Chinese readers send their eyes? In R. R. J. Hyona & H. Deubel (Eds.), *The mind's eye: Cognitive and applied aspects of eye movement research* (pp. 159-176). Amsterdam, the Netherlands: Elsevier.

Vanyukov, P. M., Warren, T., Wheeler, M. E., & Reichle, E. D. (2012). The emergence of frequency effects in eye movements. *Cognition, 123*, 185-189.

Wei, W., Li, X., & Pollastsek, A. (2013). Word properties of a fixated region affect outgoing saccade length in Chinese reading. *Vision Research, 80*, 1-6.

Williams, C. C., & Pollatsek, A. (2007). Searching for an O in an array of Cs: Eye movements track moment-to-moment processing in visual search. *Perception and Psychophysics, 69*, 372-381.

Williams, C. C., Pollatsek, A., & Reichle, E. D. (2014). Examing eye movements in visual search through clusters of objects in a circular array. *Journal of Cognitive Psychology, 26*, 1-14.

Yan, G., Tian, H., Bai, X., & Rayner, K. (2006). The effect of word and character frequency on the eye movements of Chinese readers. *British Journal of Psychology, 97*, 259-268.

Yan, M., Kliegl, R., Richter, E., Nuthmann, A., & Shu, H. (2010). Flexible saccade target selection in Chinese reading. *The Quarterly Journal of Experimental Psychology, 63*, 705-725.

Yang, H., & McConkie, G. W. (1999). Reading Chinese: Some basic eye-movement characteristics. In J. Wang, A. W. Inhoff, & H.-C. Chen (Eds.), *Reading Chinese script* (pp. 207-222). Mahwah, NJ: Erlbaum.

Zang, C., Liversedge, S.P., Bai, X., & Yan, G. (2011). Eye movements during Chinese reading. In S.P. Liversedge, I.D. Gilchrist, & S. Everling. (Eds). *Oxford Handbook on Eye Movements* (pp. 961-978). Oxford University Press.

# Footnote

1. For the purposes of brevity and to draw parallels between our paradigm and the actual reading of Chinese text (which is of theoretical interest), we will refer to the Landolt-squares used in our experiment as "characters" and to the clusters of such characters as "words" throughout the remainder of this article. (Accordingly, we will also drop the use of scare quotes.)

**Table 1**. Chi-square tests for the uniformity of all forward fixation locations as a function of word size (i.e., number of characters).

| Word Size | $\chi^2$ | $df$ | $p$ |
|---|---|---|---|
| 1-character | 0.08 | 1 | 0.778 |
| 2-character | 1.19 | 3 | 0.755 |
| 3-character | 0.92 | 5 | 0.969 |
| 4-character | 1.74 | 7 | 0.973 |

*Note*: The null hypothesis is uniformity of all forward fixation locations.

**Table 2**. LMM analyses of all forward fixation locations (aligned to the left of fixated word, in characters).

| Variables | Model | | | | Values (characters) | |
|---|---|---|---|---|---|---|
| | $b$ | $SE$ | $t$ | $p$ | Min. | Max. |
| Intercept | 0.01 | 0.07 | 0.13 | 0.894 | - | - |
| Practice (# of trials) | 0.0001 | 0.0001 | 0.40 | 0.689 | 2.46 | 2.47 |
| Word $N$-1 | | | | | | |
|   # of characters | -0.01 | 0.01 | -1.20 | 0.232 | 2.48 | 2.45 |
|   gap size (pixels) | -0.001 | 0.01 | -0.15 | 0.878 | 2.47 | 2.46 |
| Word $N$ | | | | | | |
|   # of characters | 0.99 | 0.01 | 97.27 | < 0.001 | 0.99 | 3.95 |
|   gap size (pixels) | 0.01 | 0.01 | 1.01 | 0.315 | 2.45 | 2.48 |
| Word $N$+1 | | | | | | |
|   # of characters | 0.01 | 0.01 | 0.92 | 0.358 | 2.45 | 2.48 |
|   gap size (pixels) | -0.01 | 0.01 | -1.24 | 0.214 | 2.48 | 2.45 |

*Notes*: For number of characters, min. = 1 character, max. = 4 characters; for gap size, min. = 2 pixels, max. = 8 pixels; for practice, min. = 1st trial, max. = 256th trial. Estimates of predicted variable values were calculated while fixing the values of the other variables equal to their mean values.

**Table 3**. LMM analyses of forward saccade length (in characters).

| Variables | Model | | | | Values (characters) | |
|---|---|---|---|---|---|---|
| | *b* | *SE* | *t* | *P* | Min. | Max. |
| Intercept | 2.88 | 0.19 | 15.28 | < 0.001 | - | - |
| Practice (# of trials) | 0.0001 | 0.0004 | 0.16 | 0.877 | 3.42 | 3.44 |
| Word *N*-1 | | | | | | |
|   # of characters | 0.01 | 0.01 | 2.43 | 0.015 | 3.40 | 3.44 |
|   gap size (pixels) | 0.01 | 0.003 | 1.72 | 0.085 | 3.40 | 3.43 |
| Word *N* | | | | | | |
|   # of characters | 0.10 | 0.01 | 9.88 | < 0.001 | 3.27 | 3.56 |
|   gap size (pixels) | 0.01 | 0.01 | 2.23 | 0.037 | 3.38 | 3.46 |
| Word *N*+1 | | | | | | |
|   # of characters | 0.04 | 0.01 | 5.05 | < 0.001 | 3.36 | 3.48 |
|   gap size (pixels) | 0.02 | 0.003 | 5.58 | < 0.001 | 3.37 | 3.47 |

*Notes*: For number of characters, min. = 1 character, max. = 4 characters; for gap size, min. = 2 pixels, max. = 8 pixels; for practice, min. = 1st trial, max. = 256th trial. Estimates of predicted variable values were calculated while fixing the values of the other variables equal to their mean values.

**Table 4**. LMM analyses of first-fixation durations (ms).

| Variables | Model | | | | Values (ms) | |
|---|---|---|---|---|---|---|
| | *b* | *SE* | *t* | *p* | Min. | Max. |
| Intercept | 319.70 | 10.80 | 29.61 | < 0.001 | - | - |
| Practice (# of trials) | -0.03 | 0.02 | -1.64 | 0.117 | 315 | 308 |
| Word *N*-1 | | | | | | |
|   # of characters | 0.63 | 0.49 | 1.28 | 0.200 | 310 | 312 |
|   gap size (pixels) | -0.07 | 0.34 | -0.20 | 0.847 | 312 | 311 |
| Word *N* | | | | | | |
|   # of characters | -1.35 | 0.67 | -2.01 | 0.051 | 313 | 309 |
|   gap size (pixels) | -1.16 | 0.33 | -3.50 | 0.002 | 315 | 308 |
| Word *N*+1 | | | | | | |
|   # of characters | 1.46 | 0.93 | 1.57 | 0.131 | 309 | 313 |
|   gap size (pixels) | -0.07 | 0.32 | -0.23 | 0.819 | 312 | 311 |

*Notes*: For number of characters, min. = 1 character, max. = 4 characters; for gap size, min. = 2 pixels, max. = 8 pixels; for practice, min. = 1st trial, max. = 256th trial. Estimates of predicted variable values were calculated while fixing the values of the other variables equal to their mean values.

**Table 5**. LMM analyses of gaze durations (ms).

| Variables | Model | | | | Values (ms) | |
|---|---|---|---|---|---|---|
| | *b* | *SE* | *t* | *p* | Min. | Max. |
| Intercept | 246.30 | 28.39 | 8.68 | < 0.001 | - | - |
| Practice (# of trials) | 0.04 | 0.07 | 0.59 | 0.560 | 517 | 528 |
| Word *N*-1 | | | | | | |
|   # of characters | -4.03 | 1.21 | -3.34 | < 0.001 | 529 | 517 |
|   gap size (pixels) | -1.39 | 1.02 | -1.37 | 0.187 | 527 | 519 |
| Word *N* | | | | | | |
|   # of characters | 135.70 | 10.91 | 12.44 | < 0.001 | 322 | 730 |
|   gap size (pixels) | -9.75 | 0.72 | -13.47 | < 0.001 | 552 | 494 |
| Word *N*+1 | | | | | | |
|   # of characters | -6.84 | 1.53 | -4.46 | < 0.001 | 533 | 513 |
|   gap size (pixels) | 0.55 | 0.60 | 0.92 | 0.357 | 522 | 525 |

*Notes*: For number of characters, min. = 1 character, max. = 4 characters; for gap size, min. = 2 pixels, max. = 8 pixels; for practice, min. = 1st trial, max. = 256th trial. Estimates of predicted variable values were calculated while fixing the values of the other variables equal to their mean values.

**Table 6**. LMM analyses of total-viewing times (ms).

| Variables | Model | | | | Values (ms) | |
|---|---|---|---|---|---|---|
| | *b* | *SE* | *t* | *p* | Min. | Max. |
| Intercept | 306.80 | 31.93 | 9.61 | < 0.001 | - | - |
| Practice (# of trials) | -0.08 | 0.08 | -1.00 | 0.328 | 598 | 576 |
| Word *N*-1 | | | | | | |
| # of characters | -5.06 | 1.41 | -3.59 | < 0.001 | 595 | 580 |
| gap size (pixels) | -1.66 | 1.11 | -1.49 | 0.151 | 592 | 582 |
| Word *N* | | | | | | |
| # of characters | 158.70 | 3.66 | 43.37 | < 0.001 | 349 | 825 |
| gap size (pixels) | -13.16 | 0.85 | -15.44 | < 0.001 | 627 | 548 |
| Word *N*+1 | | | | | | |
| # of characters | -6.53 | 1.41 | -4.62 | < 0.001 | 597 | 578 |
| gap size (pixels) | -0.42 | 0.70 | -0.60 | 0.548 | 589 | 586 |

*Notes*: For number of characters, min. = 1 character, max. = 4 characters; for gap size, min. = 2 pixels, max. = 8 pixels; for practice, min. = 1st trial, max. = 256th trial. Estimates of predicted variable values were calculated while fixing the values of the other variables equal to their mean values.

# Figure Caption

*Figure 1*. Examples of two Chinese sentences and their respective translations. In the first, the sequence of four underlined characters correspond to two words. In the second, the same characters correspond to a single word.

*Figure 2*. Examples of experimental materials, with the target (i.e., a square) and one of the "word" clusters being rendered in gray for illustrative purposes.

*Figure 3.* The distribution of initial fixations on words that received one or more fixations (panels a-b), of first-of-multiple fixations on words that received two or more fixations (panels c-d), of single fixations on words that received only one fixation (panels e-f), and of all forward fixations on words during first-pass scanning (panels g-h). All of the panels show the fixation-location distributions as a function of the number of characters within a word. Panels a, c, e, and g (left column) also show the fixation-location distributions as a function of the gap size within a word, whereas panels b, d, f, and g (right column) also show the fixation-location distributions as a function of practice (ordinal block number). Error bars represented the standard errors of the means.

*Figure 1.*

最终这个药铺的<u>当家做主</u>赔偿了顾客的全部损失。
Finally, the drugstore <u>owner</u> <u>decided to</u> pay for the damage of the customer.

我们要充分尊重农民<u>当家做主</u>的意愿。
We should fully respect that the peasants want to <u>make their own decisions</u>.

*Figure 2.*

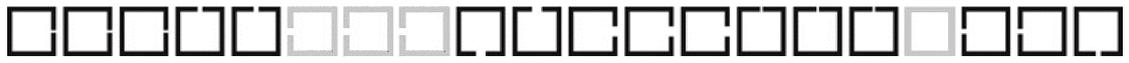

*Figure 3.*

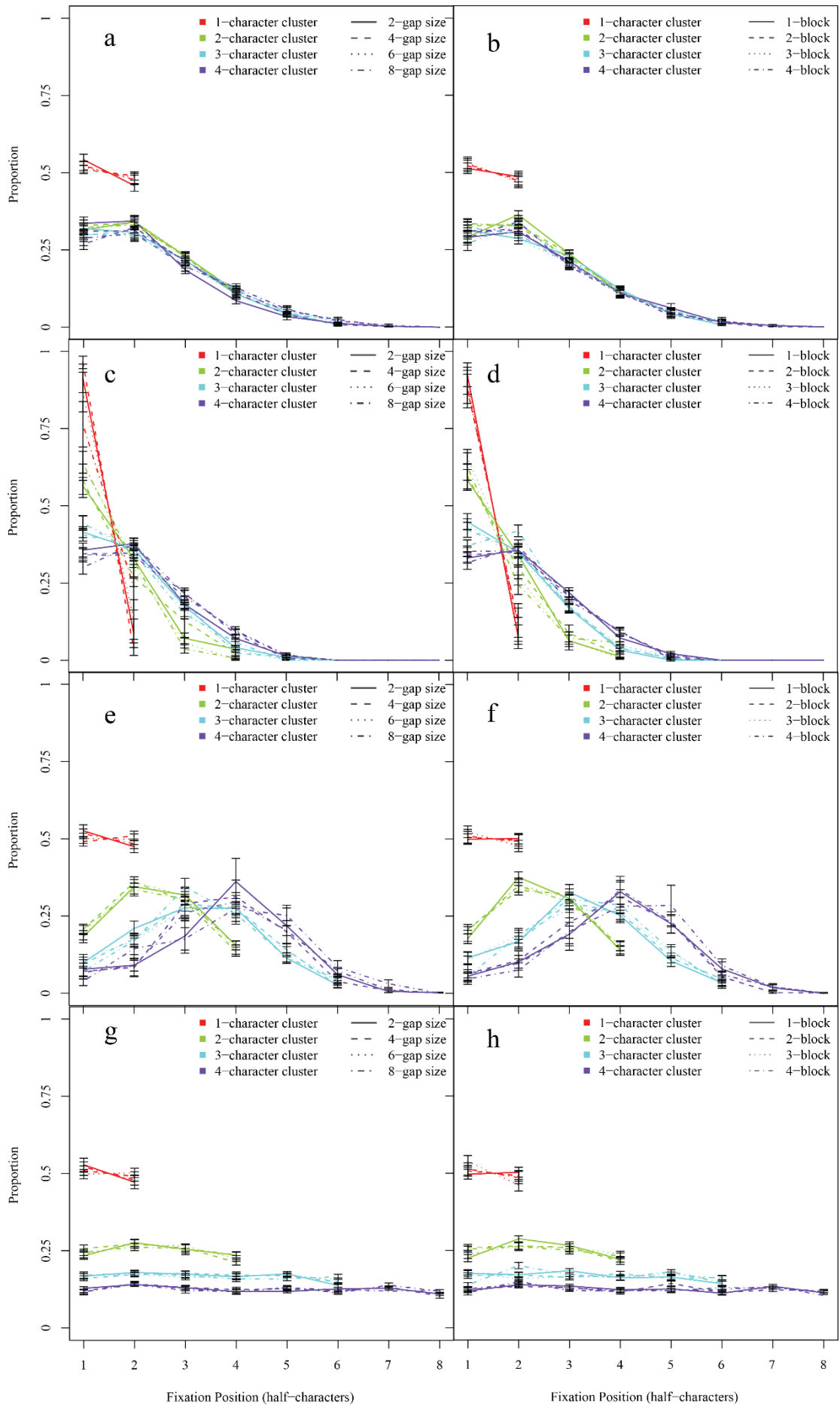